\documentclass{llncs}

\usepackage{colortbl}

\newif\ifdraft\drafttrue
\newif\ifinlineref\inlinereffalse
\newif\iffinal\finalfalse
\newif\ifextended\extendedfalse
\newif\ifdotikz\dotikzfalse

\inlinereftrue
\draftfalse
\finaltrue

\usepackage{times}
\usepackage{helvet}
\usepackage{courier}

\usepackage{ifpdf}
\ifpdf
\usepackage{microtype}
\fi

\ifdraft
\usepackage{pdfsync}
\usepackage{srcltx}
\else
\fi

\ifdraft
\usepackage{svninfo}
\svnInfo $Id: paper.tex 2969 2011-02-24 01:56:05Z drescher $
\usepackage{comments}

\else
\newcommand{\comment}[1]{}
\fi

\usepackage{paralist}
\usepackage{enumerate}

\iffinal\dotikzfalse\fi
\ifdotikz
\usepackage{tikz}
\usetikzlibrary{arrows,positioning,automata,snakes}
\pgfrealjobname{paper}
\else
\long\def\beginpgfgraphicnamed#1#2\endpgfgraphicnamed{\includegraphics{#1}}
\fi

\usepackage{amssymb}
\usepackage{amsmath}
\usepackage{multirow}
\usepackage{graphicx}
\graphicspath{{figures/}{gnuplot/}}

\newcommand{\In}[1]{\ensuremath{\mathit{In}(#1)}}
\newcommand{\Ic}[1]{\ensuremath{\mathit{IC}(#1)}}

\def\DSD{\ensuremath{\text{DSD}}}
\newcommand{\gap}[1]{\ensuremath{\mathit{GAP}(#1)}}

\newcommand\nop[1]{}
\newcommand{\card}[1]{\ensuremath{\lvert #1 \rvert}}

\newcommand{\Pol}{\ensuremath{\rm{P}}}

\newcommand{\NP}{\ensuremath{\rm{NP}}}

\newcommand{\systemname}[1]{\textsc{#1}}
\newcommand{\alphabet}{\mathcal{A}}
\newcommand{\alphabets}{\mathcal{A}}
\newcommand{\KB}[0]{\mbox{KB}}
\newcommand{\BS}[0]{\mbox{BS}}
\newcommand{\ACC}[0]{\mbox{ACC}}
\newcommand{\dneg}[0]{{\sim}}
\newcommand{\smalldneg}[0]{{\scriptstyle\sim\!}}
\newcommand{\head}[1]{\mbox{head}(#1)}
\newcommand{\body}[1]{\mbox{body}(#1)}
\newcommand{\atom}[1]{\mbox{at}(#1)}
\newcommand{\domain}[1]{\mbox{dom}(#1)}
\newcommand{\pc}[0]{\mbox{PC}}
\newcommand{\llsbc}[0]{\mbox{LL-SBC}}

\begin{document}

\abovedisplayshortskip=2pt 
\belowdisplayshortskip=2pt
\abovedisplayskip=4pt
\belowdisplayskip=4pt

\title{Symmetry Breaking for \\ Distributed Multi-Context Systems%
\iffinal%
\thanks{This research has been supported by the Austrian Science
    Fund 
    project P20841 and by the Vienna Science and Technology
    Fund 
    project ICT 08-020. 
    NICTA is funded by the Department of Broadband, Communications and the Digital Economy, and the Australian Research Council.
    }
\fi}

\author{Christian Drescher\inst{1} \and Thomas Eiter\inst{2} \and
  Michael Fink\inst{2} \and \\ Thomas Krennwallner\inst{2} \and Toby Walsh\inst{1}}

\institute{NICTA and University of New South Wales\\
\iffinal Locked Bag 6016, Sydney NSW 1466, Australia\\
\email{$\{$christian.drescher,toby.walsh$\}$@nicta.com.au}\fi
\and Institut f\"ur Informationssysteme, Technische Universit\"at Wien\\
\iffinal Favoritenstra\ss{}e\ 9-11, A-1040 Vienna, Austria\\
\email{$\{$eiter,fink,tkren$\}$@kr.tuwien.ac.at}%
\fi
}

\maketitle

\begin{abstract}
  Heterogeneous nonmonotonic multi-context systems (MCS) permit different
  logics to be used in different contexts, and link them via bridge
  rules. We investigate the role of symmetry detection and symmetry
  breaking in such systems to eliminate symmetric parts of the
  search space and, thereby, simplify the evaluation process.  We
  propose a distributed algorithm that takes a local stance, i.e.,
  computes independently the partial symmetries of a context and, in
  order to construct potential symmetries of the whole, combines them with
  those partial symmetries returned by neighbouring contexts.  We prove the
  correctness of our methods. We instantiate such symmetry
  detection and symmetry breaking in a multi-context system with contexts that use answer
  set programs, and demonstrate computational benefit on some recently proposed
  benchmarks. 
  \comment{TK: rephrased a bit \\TW: done}
\end{abstract}

\section{Introduction}
Due to the increasing application of distributed systems, 
there has been recent interest in formalisms that accommodate several, distributed knowledge bases.
Based on work by McCarthy~\cite{mccarthy78a} and Giunchiglia~\cite{giunchiglia92a}, a powerful approach is
multi-context systems (MCS; \cite{gise94a}). 
Intuitively, an MCS consists of several heterogeneous theories (the
contexts), 
which may 
\comment{TE: Small edit} use different logical languages and different inference systems, that are interlinked with a special type of rules that allow to add knowledge into a context depending on knowledge in other contexts.
MCSs have applications in various areas such as argumentation, data integration, and multi-agent systems. In the latter, each context models the beliefs of an agent while the bridge rules model an agent's perception of the environment. 
Among various proposals for MCS, the general MCS framework of Brewka and Eiter~\cite{brei07a} is of special interest, as it generalises previous approaches in contextual reasoning and allows for heterogeneous and nonmonotonic MCSs. Such a system can have different, possibly nonmonotonic logics in the different contexts, e.g., answer set programs~(ASP; \cite{baral03}), and bridge rules can use default negation to deal with incomplete information.
%

Although there has been dramatic improvements
\cite{badaeifikr10a} in the performance of distributed algorithms for
evaluating Brewka and Eiter' style nonmonotonic MCSs such as
\systemname{DMCS}~\cite{daeifikr10a}, many applications exhibit
symmetries.
For example, suppose context $C_1$ is an advanced database system which
repairs inconsistencies (e.g., from key violations in database tables),
and another context $C_2$ is accessing the repaired tables via bridge
rules. A large (exponential) number of repairs may exist, each yielding
a local model (i.e., belief set) of $C_1$; many of those models are
symmetric, thus $C_2$'s bridge rules may fire for many symmetric
repairs.
This can frustrate an evaluation algorithm as it fruitlessly explores
symmetric subspaces. Furthermore, communicating symmetric solutions from
one context to another can impede further search. If symmetries can
be identified, we can avoid redundant computation 
by pruning
parts of the search space through symmetry breaking.
However, symmetry breaking in MCSs has not been explored in any depth.

In order to deal with symmetry in MCSs, we must accomplish two tasks:
\begin{inparaenum}[(1)]
\item\label{enum:idsym} identifying symmetries and
\item\label{enum:brsym} breaking the identified symmetries.
\end{inparaenum}
We make several fundamental and foundational contributions to the study
of symmetry in MCS. 
\begin{compactitem}

\item First, we define the notion of symmetry for MCSs. This is subsequently
specialized to local symmetries and partial symmetries that capture symmetry on 
parts of an MCS. Partial symmetries can be extended to a symmetry of the whole 
system under suitable conditions which are formalized in a corresponding notion 
of join. 

\item Second, we design a distributed algorithm to identify
symmetries based on such partial symmetries. The method runs as background processes in
the contexts and communicate with each other for exchanging partial
symmetries. This algorithm computes symmetries of a general MCS based
on the partial symmetries for each individual context.
We demonstrate such symmetry detection for 
ASP contexts using automorphisms of a suitable coloured graph.
 
\item  Third, we break symmetries by extending the 
symmetry breaking methods of Crawford \emph{et al.}~\cite{crgiluro96a}
to distributed MCS. We construct
symmetry-breaking constraints~(SBCs) for a MCS that take into account beliefs 
imported from other contexts into account.
\comment{TE: TODO Stress novelty.\\
MF: Rearranged and rephrased contributions.\\TW: contributions
a little clearer and punchier} 
These constraints ensure that an evaluation engine
never visits two points in the search space that are symmetric.
For contexts other than propositional logic, distributed SBCs 
have to be expressed appropriately. Again we illustrate this in the case of ASP 
contexts and develop a logic-program encoding for distributed symmetry breaking
constraints.
 
%
\item Finally, we experimentally evaluate our approach on MCSs with ASP contexts.
In problems with large number of symmetries, we demonstrate the effectiveness
of only breaking a subset of the symmetries. 
\comment{TE: rewrite irredundant generators}
Results on MCS 
benchmarks that resemble context dependencies of realistic scenarios~\cite{badaeifikr10a} 
show that symmetry breaking yields significant improvements in runtime and compression of the solution space.
\end{compactitem}

\section{Logical Background}
We recall some basic notions of heterogeneous 
nonmonotonic multi-context systems. Following 
\cite{brei07a}, a
\emph{logic} over an alphabet $\alphabet$ is a triple~$L = (\KB, \BS,
\ACC)$, where $\KB$ is a set of
well-formed knowledge bases over~$\alphabet$, $\BS$ is a set of possible
belief sets (sets over~$\alphabet$), and $\ACC\colon \KB \to 2^{\BS}$ is
a function describing the semantics of the logic by assigning each $kb
\in \KB$ a set of acceptable sets of beliefs.
This covers many monotonic and nonmonotonic logics like
\emph{propositional logic} under the closed world assumption and
\emph{default logic}. We 
concentrate on logic programs under answer set semantics, i.e., ASP
logic~$L$. A (disjunctive) \emph{logic program} over an
alphabet~$\alphabet$ is a finite set of rules 
\begin{equation} \label{form:rule}
a_1 ; \dotsc ; a_\ell \leftarrow b_1 , \dotsc , b_j, \dneg b_{j+1} , \dotsc , \dneg b_m
\end{equation}
where $a_i, b_k \in \alphabet$ for $1 \leq i \leq \ell$, and $1 \leq k \leq m$.
A \emph{literal} is an atom~$a$ or its default negation $\dneg a$. For a
rule $r$, let
$\head{r} = \{a_1 , \dotsc , a_\ell\}$ be the \emph{head} of $r$ and
$\body{r} = \{b_1 , \dotsc , b_j, \dneg b_{j+1} , \dotsc , \dneg b_m\}$ the
\emph{body} of $r$. For an ASP logic $L$, the set of knowledge bases $\KB$ is given through the set of logic programs, the
possible belief sets~$\BS = 2^\alphabet$ contains all subsets of atoms,
and $\ACC(P)$ is the set of answer sets of a logic program~$P$. For a
detailed introduction to ASP, we refer to~\cite{baral03}. 

We now recall multi-context systems according to Brewka and Eiter~\cite{brei07a}.
%
%
A \emph{multi-context system}~$M = (C_1, \dots, C_n)$ consists of a collection of contexts~$C_i = (L_i, kb_i, br_i)$, where $L_i = (\KB_i, \BS_i, \ACC_i)$ is a logic over alphabets $\alphabet_i$, $kb_i \in \KB_i$ is a knowledge base, and $br_i$ is a set of $L_i$ \emph{bridge rules} $r$ of the form
\begin{equation} \label{form:brule}
a \leftarrow (c_1 : b_1), \dots, (c_j : b_j), \dneg (c_{j+1} : b_{j+1}),
\dots, \dneg (c_m : b_m) \enspace ,
\end{equation}
where $1 \leq c_k \leq n$, $b_k$ is an atom in $\alphabet_{c_k}$, $1 \leq k \leq m$, and $kb \cup \{a\} \in \KB_i$ for each~$kb \in \KB_i$.
%
%
We call a \emph{context atom}~$(c_k : b_k)$ or its default negation $\dneg (c_k : b_k)$ a \emph{context literal}.
Analogous to standard notions of ASP, let the atom $\head{r} = a$ be the
\emph{head} of~$r$ and $\body{r} = \{(c_1 : b_1), \dotsc, (c_j : b_j),
\dneg (c_{j+1} : b_{j+1}), \dotsc, \dneg (c_m : b_m)\}$ the \emph{body}
of $r$. For a set $S$ of context literals, define $S^{+} = \{(c~:~b)
\mid (c~:~b) \in S\}$, $S^{-} = \{(c : b) \mid \dneg (c~:~b) \in S\}$,
and for a set $S$  of context atoms, let $S\vert_c = \{ b \mid (c~:~b) \in S\}$. 
\comment{TE: Only beliefs $b$ from positive literals in $S$? \\ C: could
be extended to $(c~:~b) \in S^+ \cup S^-$, however, we below apply
$\vert_c$ to sets of context atoms only. \\ TE: Made input of $\vert_C$
a set of atoms.} 
The set of atoms occurring in a set $br_i$  of bridge rules 
is denoted by $\atom{br_i}$.
\comment{~\\~\\~\\
TE: Shortening ``atom'' to ``at'' saves space.}  W.l.o.g., we will assume
that the 
alphabets $\alphabet_i$ are pairwise disjoint
and denote their union by~$\alphabets = \bigcup_{i=1}^n \alphabet_i$.

Intuitively, context literals in bridge rules refer to information of other contexts. Bridge rules can thus modify the knowledge base, depending on what is believed or disbelieved in other contexts.
The semantics of an MCS is given by its equilibria, 
which is a collection of acceptable belief sets, one from each context,
that respect all bridge rules. More formally, for an MCS~$M = (C_1,
\dots, C_n)$ define a \emph{belief state}~$S = (S_1, \dots, S_n)$ of $M$
such that each $S_i \in \BS_i$. A bridge rule~$r$ of the
form~(\ref{form:brule}) is \emph{applicable} in 
$S$ iff
$\body{r}^+\vert_{c_k} \subseteq S_{c_k}$ and $\body{r}^-\vert_{c_k}
\cap S_{c_k} = \emptyset$ for all $1 \leq k \leq m$.
%
%
A belief state~$S = (S_1, \dots, S_n)$ of an MCS $M = (C_1, \dots, C_n)$ is an \emph{equilibrium} iff
$S_i \in \ACC_i(kb_i \cup \{\head{r} \mid r \in br_i,\ r \text{ is applicable in } S\})$ for all $1 \leq i \leq n$.

%
In practice, however, we are more interested in equilibria of a \emph{subsystem} with root context~$C_k$, e.g., when querying to a context. 
Naturally, such partial equilibria have to contain coherent information
from $C_k$ and all contexts in the import closure of $C_k$, and
therefore, are parts of potential equilibria of the whole system. We
define the \emph{import neighbourhood} of a context $C_k$ as the 
set $\In{k} = \{c \mid (c : b) \in \body{r}, r \in br_k\}$ and the
\emph{import closure} $\Ic{k}$ as the smallest set of contexts $S$ such that
\begin{inparaenum}[(1)]
\item $C_k \in S$ and 
\item $C_i \in S \text{ implies } \{ C_j \mid j \in \In{i}\} \subseteq S$.
\end{inparaenum}
Let $\varepsilon \notin \alphabets$ be a new symbol representing the value `unknown'. A \emph{partial belief state} of $M$ is a sequence $S=(S_1,\dotsc,S_n)$, such that~$S_i \in \BS_i \cup \{\varepsilon\}$ for all~$1 \leq i \leq n$.
%
%
A partial belief state $S=(S_1, \dotsc, S_n)$ of MCS $M = (C_1, \dotsc, C_n)$ w.r.t.~$C_k$ is a \emph{partial equilibrium} iff whenever $C_i \in \Ic{k}$,
$S_i \in \ACC_i(kb_i \cup \{\head{r} \mid r \in br_i,\ r \text{ is applicable in } S\})$, otherwise $S_i = \varepsilon$, for all $1 \leq i \leq n$.
%
\begin{example}\label{example:mcs}
  As a running example, consider the MCS $M = (C_1, C_2, C_3)$ with ASP logics over
  alphabets~$\alphabet_1 = \{a,b,c\}$, $\alphabet_2 = \{d,e,f,g\}$, and
  $\alphabet_3 = \{h\}$. Suppose
\[
\begin{array}{rl@{\quad}rl@{\quad}rl@{\quad}l}
kb_1 &= \left\{
\begin{array}{r@{\ \leftarrow\ }l}
c & a, b, \dneg c\\
\end{array}\right\},
&
kb_2 &= \left\{
\begin{array}{r@{\ \leftarrow\ }l}
f & d, e, \dneg g\\
g & d, e, \dneg f\\
\end{array}\right\},
&
kb_3 &= \emptyset,
\\[1em]
br_1 &= \left\{
\begin{array}{r@{\ \leftarrow\ }l}
a & \dneg (2 : d)\\
b & \dneg (2 : e)\\
\end{array}\right\},
&
br_2 &= \left\{
\begin{array}{r@{\ \leftarrow\ }l}
d & \dneg (1 : a)\\
e & \dneg (1 : b)\\
\end{array}\right\},
&
br_3 &= \left\{
\begin{array}{r@{\ \leftarrow\ }l}
h & (1 : a)
\end{array}\right\}.
\end{array}
\]
Then, $(\{b\},\{d\}, \varepsilon)$, $(\{a\},\{e\},
\varepsilon)$, $(\emptyset,\{d,e,f\}, \varepsilon)$, and
$(\emptyset,\{d,e,g\}, \varepsilon)$ are partial equilibria
w.r.t. $C_1$, and $(\{b\},\{d\}, \emptyset)$, $(\{a\},\{e\}, \{h\})$,
$(\emptyset,\{d,e,f\}, \emptyset)$, and $(\emptyset,\{d,e,g\},
\emptyset)$ are equilibria. Observe that $M$ remains invariant under a
swap of atoms~$f$ and $g$, which is what we will call a symmetry of
$M$. Furthermore, the subsystem given by~$\Ic{1} =$ $\{C_1, C_2\}$
remains invariant under a swap of atoms $f$ and $g$, and/or a
simultaneous swap of atoms $a,b$ and $d,e$, which is what we will call a
partial symmetry of~$M$ w.r.t.~$\{C_1, C_2\}$.
\end{example}
%

\section{Algebraic Background}

Intuitively, a symmetry of a discrete object is a transformation of its
components that leaves the object unchanged.  Symmetries are studied in
terms of groups. Recall that a \emph{group} is an abstract algebraic
structure~$(G,\ast)$, where $G$ is a set closed under a binary
associative operation~$\ast$ such that there is a \emph{unit} element
and every element has a unique \emph{inverse}.
Often, we abuse notation and refer to the group~$G$, rather than to the structure~$(G,\ast)$, and we denote the size of $G$ as $\card{G}$. A compact representation of a group is given through generators.
A set of group elements such that any other group element can be
expressed in terms of their product is called a \emph{generating set} or
\emph{set of generators}, and its elements are called
\emph{generators}. A generator is \emph{redundant}, if it can be
expressed in terms of other generators. A generating
set is \emph{irredundant}, if no strict subset of it is generating.
%
Such a set provides an extremely compact representation of a group. In
fact, representing a finite
group by an irredundant generating
set ensures exponential compression, 
as it contains at most~$\log_2\card{G}$ elements~\cite{almasa03a}.
\comment{TE: reformulated, pls check. \\ C: ok}

A mapping $f \colon G \to H$ between two groups~$(G,\ast)$ and $(H,
\circ)$ is a \emph{homomorphism} iff for $a,b \in G$ we have that $f(a \ast
b) = f(a) \circ f(b)$; if it has also an inverse that
is a homomorphism, $f$ is an \emph{isomorphism}, which is an
\emph{automorphism} if $G=H$.
The groups $G$ and $H$ are called \emph{isomorphic}, if there exists some isomorphism
between them. 
Any group isomorphism maps (irredundant) generating sets to
(irredundant) generating sets~\cite{almasa03a}.
%
The domain $G$ of $f$ is denoted as $\domain{f}$.
\comment{TK: definition of $\domain{}$ was missing}
In our context, 
the group of permutations is most important.
Recall that a \emph{permutation} of a set~$S$ is a bijection~$\pi \colon S \to S$.
%
%
It is well-known 
that the set of all permutations of $S$ form a group under composition, denoted
as $\Pi(S)$.


The image of $a \in S$ under a permutation~$\pi$ is denoted as~$a^\pi$,
and for vectors~$s = (a_1, a_2, \dotsc, a_k)
\in S^k$ define~$s^\pi = (a_1^\pi, a_2^\pi, \dotsc, a_k^\pi)$.
For formulas $\phi(a_1, a_2, \dotsc, a_k)$ of some logic over
alphabet~$\alphabet$ s.~t.~$S \subseteq \alphabet$ \comment{TE: This is
  a bit unclear. TK: I guess $S$ is meant as alphabet here.} define
$\phi^\pi(a_1, a_2, \dotsc, a_k) = \phi(a_1^\pi, a_2^\pi, \dotsc,
a_k^\pi)$, e.g., for a rule $r$ of form~\eqref{form:rule}, let $r^\pi$
be
 $a_1^\pi ; \dotsc ; a_\ell^\pi \leftarrow b_1^\pi , \dotsc ,
b_j^\pi, \dneg b_{j+1}^\pi , \dotsc , \dneg b_m^\pi$.
For a bridge rule~$r$ of form~\eqref{form:brule} define 
$r^\pi = a^\pi \leftarrow {(c_1 : b_1^\pi)}, \dotsc, {(c_j : b_j^\pi)}, \dneg
{(c_{j+1} : b_{j+1}^\pi)}, \dotsc,$  $\dneg {(c_m : b_m^\pi)}$. 
%
Finally, for a set $X$ (of elements or subsets from $S$, formulas, bridge rules,
etc.), define~$X^\pi = \{ x^\pi \mid x\in X \}$.
\comment{TK: definition of $kb^\pi$ and so on was missing. Generalized it.}

We will make use of the \emph{cycle notation} where a permutation is a product of disjoint cycles. A cycle~$(a_1\ a_2\ a_3\ \dotsb\ a_n)$ means that the permutation maps $a_1$ to~$a_2$, $a_2$ to~$a_3$, and so on, finally $a_n$ back to~$a_1$. An element that does not appear in any cycle is understood as being mapped to itself. 
The \emph{orbit} of~$a \in S$ under a permutation~$\pi \in \Pi(S)$ are the set of elements of~$S$ to which $a$ can be mapped by (repeatedly) applying~$\pi$.
Note that orbits define an equivalence relation on elements (sets, vectors, etc.) of~$S$.

In graph theory, the symmetries are studied in terms of graph automorphisms. We consider directed graphs~$G = (V,E)$, where~$V$ is a set of vertices and $E \subseteq V \times V$ is a set of directed edges. Intuitively, an automorphism of~$G$ is a permutation of its vertices that maps edges to edges, and non-edges to non-edges, preserving edge orientation. More formally, an \emph{automorphism} or a \emph{symmetry of $G$} is a permutation~$\pi \in \Pi(V)$ such that $(u,v)^\pi \in E$ iff $(u,v) \in E$.
An extension considers vertex colourings that are
partitionings $\rho(V) = \{V_1, V_2, \dotsc, V_k\}$ of the nodes $V$
into disjoint nonempty sets (``colours'') $V_i$. \comment{TE: add what
colouring is. \\ C: colouring is a synonym for partitioning}
Symmetries must map each vertex to a vertex with the same colour.
Formally, given a colouring of the vertices~$\rho(V) = \{V_1, V_2, \dotsc, V_k\}$, an \emph{automorphism} or a \emph{symmetry of a coloured graph} $G$ is a symmetry $\pi$ of $G$ s.t.~$\rho(V)^\pi = \rho(V)$.
%
%
The \emph{graph automorphism} problem (GAP) is to find all symmetries of
a given graph, for instance, in terms of generators. GAP is not known to
be solvable in polynomial time, and its decisional variant is known to
be within the complexity classes $\Pol$ and $\NP$, but there is strong
evidence that this problem is not $\NP$-complete (cf.~\cite{ba95a}).
Thus it is potentially easier than, for instance, deciding answer
set existence. 
\comment{TK: rephrased, NPC is not a complexity class}

\section{Symmetry in Multi-Context Systems}

We will now define our notion of a symmetry of a multi-context system.
%
In this section we consider MCS $M=(C_1,\dotsc,C_n)$ with logics~$L_i$
over alphabet~$\alphabet_i$, for $1\leq i\leq n$. \comment{TK: share $M$
  between definitions}
\begin{definition}
%
 A \emph{symmetry of~$M$} is a permutation~$\pi \in
\Pi(\alphabets)$ such that
\begin{inparaenum}[(1)]
\item\label{def:sym-alpha} $\alphabet_i^\pi = \alphabet_i$,
\item\label{def:sym-kb} $kb_i^\pi = kb_i$, and
\item\label{def:sym-br} $br_i^\pi = br_i$,
\end{inparaenum}
for $1 \leq i \leq n$.
\end{definition}
\comment{TK: as this is a key definition, we should give some intuition
  about each point.}
\comment{MF: Added paragraph below. \\ TE: Shortened.}
In this definition, 
items \eqref{def:sym-kb} and \eqref{def:sym-br} capture the intention that
symmetries are permutations of beliefs which yield identical knowledge
bases and bridge rules, respectively. 
Item~\eqref{def:sym-alpha} imposes that
symmetries do not alter
the indiviuale context languages;
there is no technical need for this, 
i.e., dropping 
\eqref{def:sym-alpha} 
would yield a more general definition of
symmetry for which our subsequent results would still hold; however the
respective additional symmetries are irrelevant from a practical point
of view and thus disregarded. For the same reason, we 
disregard permutations of the order of contexts.

Sometimes, a symmetry affects only atoms of a single context, i.e., behaves like the identity for the atoms of all other contexts. 
A symmetry~$\pi$ of~$M$ is \emph{local} for context~$C_k$ iff $a^\pi = a$ for all $a \in \domain{\pi} \setminus \alphabet_k$.
%
\begin{example}[cont'd]
  Reconsider the MCS $M = (C_1, C_2, C_3)$ from Example
  \ref{example:mcs}. Symmetries of~$M$ are given through the identity
  and $(f\ g)$, both are local for $C_2$.
\end{example}
Similar to belief states, we define the notion of partial symmetries, which are parts of potential symmetries of the system.
\begin{definition} \label{def:partial}
A permutation~$\pi$ of the elements in $S \subseteq \alphabets$ is a \emph{partial symmetry} of~$M$ w.r.t. the set of contexts~$C = \{C_{i_1}, \dotsc, C_{i_m}\}$ iff
\begin{inparaenum}[(1)]
\item $\alphabet_{i_k}\cup\atom{br_{i_k}} \subseteq S$
\item $\alphabet_{i_k}^\pi = \alphabet_{i_k}$,
\item $kb_{i_k}^\pi = kb_{i_k}$, and
\item $br_{i_k}^\pi = br_{i_k}$,
\end{inparaenum}
for all $1 \leq k \leq m$.
\end{definition}
For combining partial symmetries~$\pi$ and $\sigma$, we define their \emph{join} $\pi \bowtie \sigma$ as the permutation~$\theta$, where
\[
a^\theta = \begin{cases}
a^\pi & \text{if } a \in \domain{\pi}, \\
a^\sigma & \text{if } a \in \domain{\sigma}.
\end{cases}
\]
whenever $a^\pi = a^\sigma$ for all $a \in \domain{\pi} \cap
\domain{\sigma}$; otherwise, the join is undefined.
\comment{TE: say that join is undefined.} The \emph{join} of two sets of partial symmetries of $M$ is naturally defined as $\Pi \bowtie \Sigma = \{ \pi \bowtie \sigma \mid \pi \in \Pi,\ \sigma \in \Sigma\}$.
Note that,~$\pi \bowtie \sigma$ is void, i.e., undefined, if $\pi$ and $\sigma$ behave different for some $a \in \domain{\pi} \cap \domain{\sigma}$. Otherwise, the join is a partial symmetry of~$M$.
\begin{theorem} \label{theorem:join}
Let $M = (C_1, \dotsc, C_n)$ be an
  MCS with logics~$L_i$ over alphabet~$\alphabet_i$.
\begin{inparaenum}[(1)]
\item\label{thm:j1} Every partial symmetry of~$M$ w.r.t.~$\{C_1, \dotsc, C_n\}$ is
a symmetry of~$M$.
\item\label{thm:j2}  For every partial symmetries $\pi$ and $\sigma$ of~$M$
w.r.t.~$C_{(\pi)}= \{C_{i_1},\dotsc,C_{i_m}\}$ and $C_{(\sigma)} =
\{C_{j_1},\dotsc,C_{j_\ell}\}$,  respectively, such that $\theta = \pi \bowtie \sigma$ 
is defined, $\theta$ is a partial symmetry of~$M$ w.r.t.~$C_{(\pi)}\cup C_{(\sigma)}$.
\end{inparaenum}
\comment{TK: moved sentence from   Theorem~\ref{theorem:join} here.\\ C:
  I disagree and reverted TK's change. The second paragraph in the proof
  deals with this statement.\\
TE: Rephrased theo+proof (old version saved);}
\end{theorem}
\begin{proof}
  \eqref{thm:j1} Let $\theta$ be a partial symmetry of~$M$ w.r.t.~$\{C_1, \dotsc,
  C_n\}$. By Definition \ref{def:partial} we have $\domain{\theta}
  \subseteq \bigcup_{i=1}^n \alphabet_i = \alphabets$ (an upper bound for
  the domain of partial symmetries), and~$\alphabet_i \subseteq
  \domain{\pi}$ (lower bound for domain of partial symmetries) for $1
  \leq i \leq n$. Hence, $\theta$ is a permutation of exactly the
  elements in $\alphabets$. Given this, and since $\alphabet_i^\theta =
  \alphabet_i$, $kb_i^\theta = kb_i$ and $br_i^\theta = br_i$ holds for
  $1 \leq i \leq n$, i.e., all contexts in $M$, we have that $\theta$ is a symmetry
  of~$M$.
  \eqref{thm:j2} We check that all conditions of a partial symmetry hold for
  $\theta$. By definition of the join, $\domain{\theta} = \domain{\pi}
  \cup \domain{\sigma} \supseteq \bigcup_{k=1}^{m}(\alphabet_{i_k} \cup
  \atom{br_{i_k}}) \cup \bigcup_{k=1}^{\ell}(\alphabet_{j_k} \cup
  \atom{br_{j_k}})$. Furthermore, $\alphabet_{i_k}^\theta =
  \alphabet_{i_k}^\pi = \alphabet_{i_k}$, $kb_{i_k}^\theta =
  kb_{i_k}^\pi = kb_{i_k}$ and $br_{i_k}^\theta = br_{i_k}^\pi =
  br_{i_k}$ for all $1 \leq k \leq m$, and similarly,
  $\alphabet_{j_k}^\theta = \alphabet_{j_k}^\sigma = \alphabet_{j_k}$,
  $kb_{j_k}^\theta = kb_{j_k}^\sigma = kb_{j_k}$ and $br_{j_k}^\theta =
  br_{j_k}^\sigma = br_{j_k}$ for all $1 \leq k \leq \ell$. Hence,
  $\theta$ is a partial symmetry of~$M$
  w.r.t.~$C_{(\pi)}\cup C_{(\sigma)}$.
\qed
\end{proof}
Observe that every partial symmetry of $M$ w.r.t.~a set of
contexts $C$ is a partial symmetry of
$M$ w.r.t.~a non-empty subset of $C$; a partial symmetry can always be
written as the join of two partial symmetries.
\def\id{\mathrm{id}}
\begin{example}[cont'd]
  Reconsider $M$ from Example \ref{example:mcs}. The partial symmetries
  $\Pi$ of $M$ w.r.t. $\{C_1\}$ are given through the identity $\id$ and
  $(a\ b)\ (d\ e)$. The partial symmetries~$\Sigma$ of~$M$
  w.r.t. $\{C_2\}$ are given through $\id$, $(a\ b)\ (d\ e)\ (f\ g)$,
  and $(f\ g)$. The partial symmetries of~$M$ w.r.t.~$\{C_1, C_2\}$ are
  $\Pi \bowtie \Sigma = \Sigma$, and the partial symmetries~$\Theta$ of
  $M$ w.r.t.~$\{C_3\}$ are just $\id$ alone. The symmetries of~$M$
  are~$\Pi \bowtie \Theta = \{\id, (f\ g)\}$.
\end{example}

\section{Distributed Symmetry Detection}

In the following, we provide a distributed algorithm for detecting
symmetries of an MCS $M=(C_1, \dotsc, C_n)$. We follow Dao-Tran \emph{et
  al.}~\cite{daeifikr10a} by taking a local stance, i.e., we consider a
context~$C_k$ and those parts of the system that are in the import
closure of~$C_k$ to compute (potential) symmetries of the system. To
this end, we design an algorithm whose instances run independently at
each context node and communicate with other instances for exchanging
sets of partial symmetries. This provides a method for distributed
symmetry building.

The idea is as follows: starting from a context~$C_k$, we visit the
import closure of~$C_k$ by expanding the import neighbourhood at each
context, maintaining the set of visited contexts in a set~$H$, the
\emph{history}, until a leaf context is reached, or a cycle is detected
by noticing the presence of a neighbour context in~$H$. A leaf
context~$C_i$ simply computes all partial symmetries of~$M$ w.r.t.~$\{
C_i \}$. Then, it returns the results to its parent (the invoking context),
for instance, in form of permutation cycles.
The results of intermediate contexts~$C_i$ are partial symmetries of~$M$ w.r.t.~$\{ C_i\}$, which can be joined, i.e., consistently combined, with partial symmetries from their neighbours, and resulting in partial symmetries of~$M$ w.r.t.~$\Ic{i}$. In particular, the starting context~$C_k$ returns its partial symmetries joined with the results from its neighbours, as a final result.
\begin{figure}[t]
\[
\begin{array}{ll}
$\textbf{Algorithm}:$ & \DSD(H) $ at context $C_k\\
$\textbf{Input}:$ & $Visited contexts $H$.$\\
$\textbf{Data}:$ & $Cache $c(k)$.$\\
$\textbf{Output}:$ & $The set of accumulated partial symmetries $\Pi$.$\\
\end{array}
\]
\[
\begin{array}{l}
$\textbf{if} $ c(k) $ is not initialised \textbf{then} 
 $c(k) \leftarrow \text{LSD}(C_k)$;$\\
H \leftarrow H \cup \{ k \}$;$\\
\Pi \leftarrow c(k)$;$\\
$\textbf{foreach} $i \in \In{k} \setminus H$ \textbf{do} 
$\Pi\leftarrow\Pi \bowtie C_i.\DSD(H)$;$\\
$\textbf{return} $\Pi$;$
\end{array}
\]
\vspace{-1.5em}
\caption{The distributed symmetry detection algorithm. \label{alg:dsd}}
\vspace{-1.5em}
\end{figure}
We assume that each context~$C_k$ has a background process 
that waits for incoming requests with history~$H$, upon which it
starts the computation outlined in our algorithm shown in
Fig.~\ref{alg:dsd}. We write $C_i.\DSD(H)$ to specify that we send~$H$
to the process at context~$C_i$ and wait for its return message. This
process also serves the purpose of keeping the cache~$c(k)$
persistent.
\comment{TE: What precisely do you want to say with ``keeping persistent''?}  We use the primitive $\text{LSD}(C_k)$ which computes all
partial symmetries of $M$ w.r.t.  $\{C_k\}$ over $\alphabet_k \cup
\atom{br_k}$.

Our algorithm proceeds in the following way:
\begin{compactenum}
\item Check the cache for partial symmetries of $M$ w.r.t. $\{ C_k \}$;
\item if imports from neighbour contexts are needed, then request
  partial symmetries from all neighbours and join them (previously
  visited contexts excluded). This can be performed in parallel. Also,
  partial symmetries can be joined in the order neighbouring contexts do
  answer; and
\item return partial symmetries of $M$ w.r.t. $\Ic{k}$.
\end{compactenum}
Correctness of our approach 
hold by the following result.
\begin{theorem} \label{theorem:dsd}
  Let $M=(C_1,\dotsc,C_n)$ be an MCS and $C_k$ be a context in
  $M$. Then,~$\pi \in C_k.\DSD(\emptyset)$ iff $\pi$ is a partial
  symmetry of~$M$ w.r.t.~$\Ic{k}$.
\comment{TK: specify $C_k$}
\end{theorem}
\begin{proof}[sketch]
\noindent
($\Rightarrow$) We prove soundness, i.e., if $\pi \in
C_k.\DSD(\emptyset)$ then $\pi$ is a partial symmetry of~$M$
w.r.t.~$\Ic{k}$. We proceed by structural induction on the topology of an MCS, and start with acyclic MCS~$M$.
Base case: $C_k$ is a leaf with $br_k = \emptyset$ and $\In{k} = \emptyset$. By assumption, $\text{LSD}(C_k)$ computes all partial symmetries of~$M$ w.r.t.~$\{C_k\}$, i.e., $c(k) \leftarrow \text{LSD}(C_k)$ in the algorithm in Fig.~\ref{alg:dsd}.
Induction step: 
for non-leaf $C_k$, suppose $\In{k} =
\{i_1, \dotsc, i_m\}$ and $\Pi_{k} = \text{LSD}(C_k)$,
%
$\Pi_{i_j} = C_{i_j}.\DSD(H \cup \{k\})$ for~$1\leq j\leq m$.
By Theorem \ref{theorem:join}, 
$\Pi = \Pi_k \,{\bowtie}\,
\Pi_{i_1} \,{\bowtie}\, \dotsb \,{\bowtie}\, \Pi_{i_m}$, as computed by
$\Pi\leftarrow\Pi \bowtie C_i.\DSD(H)$ in the loop of the algorithm in
Fig.~\ref{alg:dsd}, 
consists of partial symmetries of~$M$ w.r.t.~$\Ic{k}$.

The proof for cyclic $M$
is similar. In a run we eventually end up in~$C_i$ such that $i \in H$ again. In that case, calling $C_i.\DSD(H)$ is discarded, which breaks the cycle. However, partial symmetries excluding $C_i$ are propagated through the system to the calling $C_i$ which combines the intermediate results with partial symmetries of~$M$ w.r.t.~$\{C_i\}$.

\noindent
($\Leftarrow$) We give now a proof sketch for completeness. Let $\pi$ be a partial symmetry of~$M$ w.r.t.~$\Ic{k}$. We show $\pi \in C_k.\DSD(\emptyset)$. The proof idea is as follows: we proceed as in the soundness part by structural induction on the topology of~$M$, and in the base case for a leaf context~$C_k$, by assumption, we get that $\text{LSD}(C_k)$ returns all partial symmetries of~$M$ w.r.t.~$\{C_k\}$, i.e., all partial symmetries of~$M$ w.r.t.~$\Ic{k}$. For the induction step, we verify straightforward that $\pi$ being a partial symmetry of~$M$ w.r.t.~$\Ic{k}$ implies $\pi$ being a partial symmetry of~$M$ w.r.t.~$\Ic{i}$ for all $i \in \In{k}$.
\qed
\end{proof}

\section{Symmetry Detection via Graph Automorphism}
The primitive $\text{LSD}(C_i)$ for detecting partial symmetries of an MCS $M = (C_1,
\dotsc, C_n)$ w.r.t.~$\{C_i\}$ using logic $L_i$ has to be defined for
every logic $L_i$ anew. As an example, our approach for detecting
partial symmetries of $M$ w.r.t. an ASP context $C_i$ is through
reduction to, and solution of, an associated graph automorphism problem.\comment{TK: GAP was
  used without definition.\\C: GAP is now defined earlier.}  The graph~$\gap{C_i}$ is constructed as follows:
\begin{compactenum}
\item\label{it:gap:1} Every atom that occurs in $kb_i \cup br_i$ (every context atom $(c : b)$ in $br_i$, respectively) is represented by two vertices of colour~$i$ ($c$, respectively) and $n+1$ that correspond to the positive and negative literals.
\item\label{it:gap:2} Every rule (every bridge rule, respectively) is represented by a \emph{body vertex} of colour~$n+2$ ($n+3$, respectively), a set of directed edges that connect the vertices of the literals (context literals, respectively) that appear in the rule's body to its body vertex, and a set of directed edges that connect the body vertex to the vertices of the atoms that appear in the head of the rule.
\item
To properly respect negation,
\comment{TE: Would say: To properly respect negation. Implemented}
that is, an atom $a$ maps to $b$ if and only if $\dneg a$ maps to $\dneg b$ for any atoms~$a$ and $b$, vertices of opposite (context) literals are mated by a directed edge from the positive (context) literal to the negative (context) literal.
\end{compactenum}
\begin{example}[cont'd]
Reconsider MCS $M$ from Example
\ref{example:mcs}. Fig.~\ref{fig:mcs} illustrates $\gap{C_2}$, where different shapes and tones represent different colours.
\end{example}
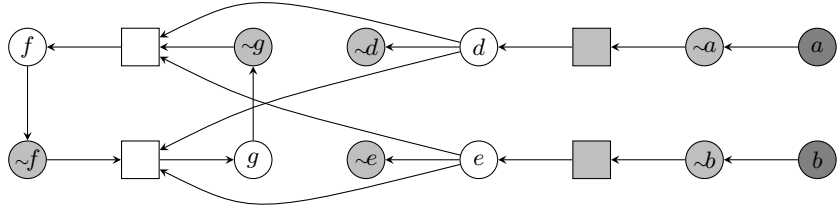
\begin{figure}[t]
\centering
\beginpgfgraphicnamed{gap-c2}
\begin{tikzpicture}
	[circle, inner sep=0pt, minimum size=5.0mm, >=stealth]
	\node (lnf) at (0.0,0.0) [draw, fill=lightgray] {$\smalldneg f$};
	\node (lf)  at (0.0,1.5) [draw] {$f$};
	\node (b2)  at (1.5,0.0) [rectangle, draw] {};
	\node (b1)  at (1.5,1.5) [rectangle, draw] {};
	\node (lg)  at (3.0,0.0) [draw] {$g$};
	\node (lng) at (3.0,1.5) [draw, fill=lightgray] {};
	\node       at (3.0,1.5) [] {$\smalldneg g$};
	\node (lnd) at (4.5,1.5) [draw, fill=lightgray] {$\smalldneg d$};
	\node (ld)  at (6.0,1.5) [draw] {$d$};
	\node (b3)  at (7.5,1.5) [rectangle, draw, fill=lightgray] {};
	\node (lna) at (9.0,1.5) [draw, fill=lightgray] {$\smalldneg a$};
	\node (la)  at (10.5,1.5) [draw, fill=gray] {$a$};
	\node (lne) at (4.5,0.0) [draw, fill=lightgray] {$\smalldneg e$};
	\node (le)  at (6.0,0.0) [draw] {$e$};
	\node (b4)  at (7.5,0.0) [rectangle, draw, fill=lightgray] {};
	\node (lnb) at (9.0,0.0) [draw, fill=lightgray] {$\smalldneg b$};
	\node (lb)  at (10.5,0.0) [draw, fill=gray] {$b$};
	\draw [->] (lnf) -- (b2);
	\draw [->] (b2) -- (lg);
	\draw [->] (lng) -- (b1);
	\draw [->] (b1) -- (lf);
	\draw [->] (lf) -- (lnf);
	\draw [->] (lg) -- (lng);
	\draw [->] (ld) -- (lnd);
	\draw [->] (le) -- (lne);
	\draw [->] (la) -- (lna);
	\draw [->] (lb) -- (lnb);
	\draw [->] (lna) -- (b3);
	\draw [->] (lnb) -- (b4);
	\draw [->] (b3) -- (ld);
	\draw [->] (b4) -- (le);
	\draw [->] (ld) .. controls (3.0,2.25)  .. (b1);
	\draw [->] (le) .. controls (3.0,0.75) .. (b1);
	\draw [->] (ld) .. controls (3.0,0.75) .. (b2);
	\draw [->] (le) .. controls (3.0,-0.75) .. (b2);
\end{tikzpicture}
\endpgfgraphicnamed
\vspace{-1.0em}
\caption{GAP reduction of context $C_2$ from Example \ref{example:mcs}. \label{fig:mcs}}
\vspace{-1.5em}
\end{figure}
Symmetries of $\gap{C_i}$ correspond precisely to the partial symmetries of~$M$ w.r.t.~$\{C_i\}$.
\begin{theorem} \label{theorem:gap}
Let $M=(C_1,\dotsc,C_n)$ be an MCS with ASP context $C_i$. 
The partial symmetries of $M$ w.r.t.~$\{C_i\}$ correspond one-to-one to the symmetries of $\gap{C_i}$.
\end{theorem}
\begin{proof}
The proof for logic programs is shown in~\cite{drtiwa10b}. Therefore we only provide arguments regarding bridge rules and context atoms.
%
($\Rightarrow$) A partial symmetry of~$M$ w.r.t.~$\{C_i\}$ will map
context atoms to context atoms of the same context. Since they have the
same colour, the symmetry is preserved for corresponding vertices and
consistency edges. The same applies to body vertices and edges representing bridge rules, since the body vertices have incoming edges from context literal vertices with their respective colour only, and vertices of the same colour are mapped one to another. Thus, a consistent mapping of atoms in $C_k$, when carried over to the graph, must preserve symmetry.
%
($\Leftarrow$) We now show that every symmetry in the graph corresponds
to a partial symmetries of~$M$ w.r.t. $\{C_i\}$. Recall that we use one
colour for positive context literals from each context, one for negative
context literals from each context, and one for bodies. Hence, a graph
symmetry must map
\begin{inparaenum}[(1)]
\item positive context literal vertices to other such from the same
  context, negative literal vertices to negative literal vertices from
  the same context, and body vertices to body vertices, and
\item the body edges of a vertex to body edges of its mate. 
\end{inparaenum}
This is consistent with partial symmetries of~$M$ w.r.t. $\{C_i\}$ mapping context atoms to context atoms, and bodies to bodies, i.e., bridge rules to bridge rules.
\qed
\end{proof}
\begin{corollary} \label{theorem:isomorph}
Let $M=(C_1,\dotsc,C_n)$ be an MCS with ASP context $C_i$.
The partial symmetry group of $M$ w.r.t.~$\{C_i\}$ and the symmetry group of $\gap{C_i}$ are isomorphic. Furthermore, sets of partial symmetry generators of $M$ w.r.t.~$\{C_i\}$ correspond one-to-one to sets of symmetry generators of $\gap{C_i}$.
\comment{TK: nop'd proof to save space}
\end{corollary}
\nop{
\begin{proof}
We easily verify
that the one-to-one mapping from Theorem
\ref{theorem:gap} is a homomorphism, and recall that an isomorphism maps
generating sets to generating sets.
\qed
\end{proof}
}
To detect local symmetries only, we further modify our approach by assigning a unique colour to each context atom and each atom that is referenced in other contexts, i.e., context atoms cannot be mapped.

With reference to related work (cf. \cite{almasa03a,drtiwa10b}), we stretch that the detection of symmetries through reduction to graph automorphism is computationally quite feasible, i.e., the overhead cost in situations that do not have symmetries is negligible.

\section{Distributed Symmetry-breaking Constraints}
Recall that a (partial) symmetry of an MCS defines equivalence classes on its (partial) equilibria through orbits. Symmetry breaking amounts to selecting some representatives from every equivalence class and formulating conditions, composed into a (distributed) symmetry-breaking constraint~(SBC), that is only satisfied on those representatives.
A \emph{full} SBC selects exactly one representative from each orbit,
otherwise we call an SBC \emph{partial}.
\comment{TE: perhaps better and a \emph{partial} SBC at least one. Total
will be regarded as special partial, I guess} The most common approach is to
order all elements from the solution space lexicographically, and to
select the lexicographically smallest element, the \emph{lex-leader},
from each orbit as its representative (see, for instance,
\cite{almasa03a,crgiluro96a}). A \emph{lex-leader symmetry-breaking
  constraint} (LL-SBC) is an SBC that is satisfied only on the
lex-leaders of orbits.
Given an MCS $M = (C_1,\dotsc,C_n)$ with logics $L_i$ over alphabet
$\alphabet_i$, we will assume a total ordering~$<_\alphabets$ on the
atoms $a_1, a_2, \dotsc, a_m$\comment{C: new index $m$ to avoid confusion with contexts} in $\alphabets$ and consider the induced
lexicographic ordering on the (partial) belief states. Following
\cite{crgiluro96a}, 
we obtain an LL-SBC by encoding a (distributed) \emph{permutation
  constraint} (PC) for every permutation $\pi$, where
%
\centerline{
$
\pc(\pi) = \bigwedge_{1 \leq i \leq m} \left\lbrack \bigwedge_{1 \leq j \leq i-1} (a_j = a_j^\pi) \right\rbrack \rightarrow(a_i \leq a_i^\pi).
$}
%
By \emph{chaining}, which uses
atoms~$c_{\pi,i}$, $1 \,{<}\, i \,{\leq}\, m{+}1$ (which informally express that for some 
$i \leq j \leq m$ the implication fails if it did not for some $j<i$),
\comment{TE: Added informal explanation, check}
we achieve a representation that is linear in the number of atoms~\cite{almasa03a}:
\[
\begin{array}{rcl}
\pc(\pi) & = & (a_1 \leq a_1^\pi) \land \neg c_{\pi,2},\\
\neg c_{\pi,i} & \leftrightarrow &((a_{i-1} \geq a_{i-1}^\pi) \rightarrow (a_i \leq a_i^\pi) \land c_{\pi,i+1})\qquad 1 < i  \leq m,\\
\neg c_{\pi,m+1} & \leftrightarrow & \top.
\end{array}
\]
\comment{TE: $\bot$ should be $\top$? \\C: yes, $\top$, of course } 
In order to distribute the $\pc$
formula in $M$, given the total ordering~$<_\alphabets$, we define (the
truth of) atoms~$c_{\pi,i}$ in the contexts $C_k$ such that $a_{i-1} \in
\alphabet_k$. 
Observe that, for each subformula, the atoms $a_i$, $a_i^\pi$ and
$c_{\pi,i+1}$ might be defined in a different context $j$, and their
truth value has to be imported via bridge rules.
We thus introduce
auxiliary atoms~$a'_i$,~$a_i^{'\pi}$, and $c'_{\pi,i+1}$ in
$C_k$ that resemble the truth of~$a_i$, $a_i^\pi$, and $c_{\pi,i+1}$,
respectively. Then we distribute $\pc(\pi)$ to each context $C_k$ for each $1
\leq k \leq n$ as follows:
\[
\begin{array}{rcl@{\, \, \text{if }}l}
\pc(\pi) & = & (a_1 \leq a_1^\pi) \land \neg c_{\pi,2} & a_1 \in \alphabet_k,\\
\neg c_{\pi,i} & \leftrightarrow &((a_{i-1} \geq a_{i-1}^\pi) \rightarrow (a_i \leq a_i^\pi) \land \neg c_{\pi,i+1}) & a_{i-1},a_i \in \alphabet_k, \\
\neg c_{\pi,i} & \leftrightarrow &((a_{i-1} \geq a_{i-1}^\pi) \rightarrow (a'_i \leq a_i^{'\pi}) \land \neg c'_{\pi,i+1}) & a_{i-1} \in \alphabet_k, a_i \in \alphabet_j, j \neq k, \\ \vspace {0.5em}
\neg c_{\pi,m+1} & \leftrightarrow & \top & a_m \in \alphabet_k, \\
a'_i &\leftarrow & (j : a_i) & a_{i-1} \in \alphabet_k, a_i \in \alphabet_j, j \neq k, \\
a_i^{'\pi} &\leftarrow & (j : a_i^\pi) & a_{i-1} \in \alphabet_k, a_i \in \alphabet_j, j \neq k, \\
c'_{\pi,i+1} &\leftarrow & (j : c_{\pi,i+1}) & a_{i-1} \in \alphabet_k, a_i \in \alphabet_j, j \neq k.
\end{array}
\]
\comment{TE: $\bot$ should be $\top$?}  
The distributed $\pc$ can be
adjusted to other logics as well. 
Exploiting detected symmetries has been studied, e.g., in the context of SAT~\cite{almasa03a,crgiluro96a}, planning~\cite{folo99a}, and constraint programming~\cite{puget05a}.
\comment{TK: we should give a statement how one in principle can define
  PCs for arbitrary logics, or for certain classes of logics. PCs for
  propositional logics have already been investigated, so what do we
  need to incorporate the bridge rule business? Then we can continue to
  give a characterisation of PCs for ASP context as a showcase.}%
For an ASP context $C_k$, we can express the distributed $\pc$ as follows: 

\noindent
\begin{minipage}{.5\textwidth}
\noindent\begin{tabular}{r@{$\ \leftarrow\ $}l@{}c@{}l}
& $a_1, \dneg a_1^\pi$ & \multirow{2}{*}{$\left.\begin{array}{l@{}@{}}\\
      \\ \end{array}\right\}$} & \multirow{2}{*}{
\begin{tabular}{l} 
if $a_1 \in \alphabet_k$;
\end{tabular}
} \\
& $c_{\pi,2}$ & & \\ \noalign{\vspace {0.5em}}
$c_{\pi,i}$ & $ a_{i-1}, a_i, \dneg a_i^\pi$ &
\multirow{4}{*}{$\left.\begin{array}{l@{}@{}}\\ \\ \\ \\ \end{array}\right\}$}
& \multirow{4}{*}{
\begin{tabular}{l}
if $a_i \in \alphabet_k$, \\
$a_{i-1}\in \alphabet_k$;
\end{tabular}} \\
$c_{\pi,i}$ & $\dneg a_{i-1}^\pi, a_i, \dneg a_i^\pi$ & & \\
$c_{\pi,i}$ & $a_{i-1}, c_{\pi,i+1}$ & & \\
$c_{\pi,i}$ & $\dneg a_{i-1}^\pi, c_{\pi,i+1}$ & & \\ \noalign{\vspace {0.5em}}
\end{tabular}
\end{minipage}
\!\!\!
\begin{minipage}{.5\textwidth}
%
\begin{tabular}{@{}r@{$\ \leftarrow\ $}l@{}c@{}l}
$c_{\pi,i}$ & $a_{i-1}, a'_i, \dneg {a'_i}^\pi$ &
\multirow{4}{*}{$\left.\begin{array}{@{}l@{}}\\ \\ \\ \\ \end{array}\right\}$}
& \multirow{4}{*}{
\begin{tabular}{l}
if $a_i \in \alphabet_{j}$, \\
$a_{i-1} \in \alphabet_k$, \\
$j \neq k$;
\end{tabular}} \\
$c_{\pi,i}$ & $\dneg a_{i-1}^\pi, a'_i, \dneg {a'_i}^\pi$ & & \\
$c_{\pi,i}$ & $a_{i-1}, c'_{\pi,i+1}$ & & \\
$c_{\pi,i}$ & $\dneg a_{i-1}^\pi, c'_{\pi,i+1}$ & & \\ \noalign{\vspace {0.5em}}
$a'_i$ & $(j : a_i)$ & \multirow{3}{*}{$\left.\begin{array}{@{}l@{}}\\ \\ \\ \end{array}\right\}$}
& \multirow{3}{*}{
\begin{tabular}{l}
if $a_i \in \alphabet_{j}$, \\
$a_{i-1} \in \alphabet_k$, \\
$j \neq k$;
\end{tabular}} \\
$a_i^{'\pi}$ & $(j : a_i^\pi)$ & \\
$c'_{\pi,i+1}$ & $(j : c_{\pi,i+1})$ & & \\ \noalign{\vspace {0.5em}}
\end{tabular}
\end{minipage}
\comment{TE: $c_{\pi,m+1}\leftarrow$ s.b.  $\leftarrow c_{\pi,m+1}$  ?\\C: nope, we simply delete that rule. Clark's completion sets $c_{\pi,m+1} \rightarrow \bot$ which is equivalent to $\neg c_{\pi,m+1} \leftrightarrow \top$.
} 

\noindent
Here, $c_{\pi,i}$ is defined from $\neg c_{\pi,i} \leftrightarrow
(\alpha \rightarrow \beta \land \neg c_{\pi,i+1})$ via $c_{\pi,i} \leftrightarrow
(\alpha \land \neg \beta \lor \alpha \land c_{\pi,i+1})$ exploiting
Clark completion and splitting $\alpha = a_{i-1}\leq a^\pi_{i-1}$ into
the (overlapping) cases where $a_{i-1}$ is true and $a^\pi_{i-1}$ is false.
We collect the newly introduced formulas in $kb_{k,\pi}$ and bridge rules in $br_{k,\pi}$ for each $1 \leq k \leq n$.
The following correctness result can be shown, generalizing a similar
result for ASP programs in \cite{drtiwa10b}.

%
\comment{TK: basic intuition about the rules is missing,
  esp. $b_{\pi,i-1} \gets (k : a_i), \dneg (k : {a_i}^\pi)$ and the
  relationship to the whole MCS.\\ C: any better now? please check and revise}%
%
\comment{TK: \cite{drtiwa10b} does not contain a formal result (at least
  the version on arxiv). Is there something else available?\\C: \cite{drtiwa10b} refers to an extended version submitted to AI Communications}%
\comment{C: correctness of $\pc$ has been shown in \cite{almasa03a,crgiluro96a}, the rest follows \emph{by construction}}%
\begin{theorem}\label{thm:sbc}
Let $\pi$ be a (partial) symmetry of an MCS $M=(C_1,\ldots,C_n)$ with
ASP contexts $C_i$. A (partial) equilibrium of $M$ satisfies $\pc(\pi)$ iff it is a (partial) equilibrium of $M(\pi)=(C_1(\pi),\dotsc,C_n(\pi))$, where $C_k(\pi)$ extends $C_k$ by $kb_k(\pi) = kb_k \cup kb_{k,\pi}$ and $br_k(\pi) = br_k \cup br_{k,\pi}$.
\end{theorem}
\comment{TE: Added sentence to clarify outreach} This result generalizes to MCS having contexts $C_i$ with (possibly heterogeneous)
logics~$L_i$ that permit to encode PC via additional formulas in the
knowledge base $kb_i$.

\begin{example}[cont'd]
Reconsider $M$ from Example~\ref{example:mcs}. Given the ordering $a
<_\alphabet b <_\alphabet d <_\alphabet e$, the permutation constraint
to break the partial symmetry 
 $\pi = (a\ b)\ (d\ e)$ is: 
%

\smallskip
\begin{tabular}{r@{$\ \leftarrow\ $}lcl}
& $a, \dneg b$ \\
& $c_{\pi,2}$ 
\end{tabular}
\quad
\begin{tabular}{r@{$\ \leftarrow\ $}lcl}
%
$c_{\pi,2}$ & $b, d', \dneg e'$ & \multirow{4}{*}{$\left.\begin{array}{l} \\ \\ \\ \\ \end{array}\right\}$} & \multirow{4}{*}{$kb_{1,\pi}$, }\\
%
$c_{\pi,2}$ & $\dneg a, d', \dneg e'$ \\
$c_{\pi,2}$ & $b, c'_{\pi,3}$ \\
$c_{\pi,2}$ & $\dneg a, c'_{\pi,3}$ 
\end{tabular}
\qquad
\begin{tabular}{r@{$\ \leftarrow\ $}lcl}
$d'$ & $(2 : d)$ & \multirow{3}{*}{$\left.\begin{array}{l} \\ \\ \\ \end{array}\right\}$} & \multirow{3}{*}{$br_{1,\pi}$, and}\\
$e'$ & $(2 : e)$ \\
$c'_{\pi,3}$ & $(2 : c_{\pi,3})$
\end{tabular}
\smallskip

\noindent
$kb_{2,\pi} \,{=}\, br_{2,\pi} \,{=}\, \emptyset$.
One can check that $(\{b\},\{e\},\varepsilon)$, $(\emptyset,\{d,e,f\},\varepsilon)$, and $(\emptyset,\{d,e,g\},\varepsilon)$ are partial equilibria of $M(\pi)$ w.r.t $C_1$, and $(\{a\},\{d\},\varepsilon)$ is not (cf. Example~\ref{example:mcs}) since $(\{a\},\{d\},\varepsilon) <_\alphabet (\{b\},\{e\},\varepsilon)$.
\comment{TK: explicit ordering of the alphabet and more intuition is
  missing. moreover, the equilibria should be listed and their
  relationship to the equilibria of Example~\ref{example:mcs}
  discussed.}
\end{example}
The LL-SBC that breaks every (partial) symmetry in an MCS, denoted $\llsbc(\Pi)$, can now be constructed by conjoining all of its permutation constraints~\cite{crgiluro96a}.
We can add $\llsbc(\Pi)$ to $M$, say
$M(\Pi)=(C_1(\Pi),\dotsc,C_n(\Pi))$, where $C_k(\Pi)$ extends $C_k$
by~$kb_k(\Pi) = kb_k \cup \bigcup_{\pi \in \Pi} kb_k(\pi)$ and
$br_k(\Pi) = br_k \cup \bigcup_{\pi \in \Pi} br_k(\pi)$.

Breaking all symmetries may not speed up search because there are often
exponentially many of them. A better trade-off may be provided by
breaking enough symmetries~\cite{crgiluro96a}. We explore partial SBCs,
i.e., we do not require that SBCs are satisfied by lex-leading
assignments only (but we still require that all lex-leaders satisfy
SBCs). Irredundant generators are good candidates because they cannot be
expressed in terms of each other, and implicitly represent all
symmetries.
\comment{we should write here a bit more and make it more formal and
  clearer, as the experiments are based on this.}
Hence, breaking all symmetry \comment{TE: symmetries? \\ C: don't know,
this was suggested by Toby \\ TE: OK.}
in a generating set can eliminate all problem symmetries.

\section{Experiments}

We present some results on breaking local symmetries in terms of
irredundant generators for distributed nonmonotonic 
MCS with ASP logics. Experiments consider the \systemname{dmcs}
system~\cite{daeifikr10a} and its optimized version
\systemname{dmcsopt}~\cite{badaeifikr10a}.
Both systems are using the ASP solver
\systemname{clasp}~\cite{gekanesc07b} as their core reasoning
engine. However, in contrast to \systemname{dmcs}, \systemname{dmcsopt}
exploits the topology of an MCS, that is the graph where contexts are
nodes and import relations define edges, using decomposition techniques
and minimises communication between contexts by projecting partial
belief states to relevant atoms.
We 
compare the average response time and the number of solutions under
symmetry breaking, denoted as \systemname{dmcs}$^\pi$ and
\systemname{dmcsopt}$^\pi$, respectively, on benchmarks versus direct application of the respective systems.
All tests were run on a 2$\times$1.80~GHz PC under Linux, where each run
was limited to 
180  seconds.
Our benchmarks stem from \cite{badaeifikr10a} and include random MCSs
with various fixed topologies that should resemble the context
dependencies of realistic scenarios. 
Experiments consider
MCS instances with ordinary~(D) and zig-zag~(Z) diamond stack, house
stack (H), and ring (R). A diamond stack combines multiple diamonds in a
row, where ordinary diamonds (in contrast to zig-zag diamonds) have no
connection between the 
2 
middle contexts. A house consists of 5 nodes
with~6 edges such that the ridge context has directed edges to the 2 
middle contexts, which form with the 
2 base contexts a cycle with 4
edges. House stacks are subsequently built 
using the basement nodes as ridges for the next houses.
\begin{table}[t]
\centering
\caption{Completed runs (10 random instances each): avg. running time (secs)
 vs. timeouts 
 \label{tab:dmcs}}
\begin{tabular}{|lc|cccccccc|}
\hline
& & \multicolumn{2}{c}{\systemname{dmcs}} & \multicolumn{2}{c}{\systemname{dmcs}$^\pi$} & \multicolumn{2}{c}{\systemname{dmcsopt}} & \multicolumn{2}{c|}{\systemname{dmcsopt}$^\pi$} \\
& $n$ & time & \#t.out & time & \#t.out & time & \#t.out & time & \#t.out \\ \hline \noalign{\smallskip} \hline
D
  &  10 &  1.90 &    &  0.46 &    &  0.54 &   &  0.35 &  \\
  &  13 & 62.12 &  4 & 32.21 &  2 &  1.38 &   &  0.98 &  \\
  &  25 &  ---  & 10 &  ---  & 10 & 16.12 &   & 11.72 &  \\
  &  31 &  ---  & 10 &  ---  & 10 & 84.02 & 1 & 58.95 &  \\ \hline
H
  &   9 &  7.54 &    &  1.89 &    &  0.33 &   &  0.20 &  \\
  &  13 & 88.85 &  6 & 63.98 &  2 &  0.60 &   &  0.35 &  \\
  &  41 &  ---  & 10 &  ---  & 10 &  1.38 &   &  0.95 &  \\
  & 101 &  ---  & 10 &  ---  & 10 &  5.48 &   &  3.58 &  \\ \hline
R
  &  10 &  0.36 &    &  0.26 &    &  0.15 &   &  0.12 &  \\
  &  13 & 22.41 &  1 &  5.11 &    &  0.19 &   &  0.16 &  \\ \hline
Z
  &  10 &  6.80 &    &  3.24 &    &  0.62 &   &  0.37 &  \\
  &  13 & 57.58 &  3 & 42.93 &  3 &  1.03 &   &  0.68 &  \\
  &  70 &  ---  & 10 &  ---  & 10 & 18.87 &   &  9.98 &  \\
  & 151 &  ---  & 10 &  ---  & 10 & 51.10 &   & 30.15 &  \\
\hline
\end{tabular}
\vspace*{-2em}
\end{table}

\comment{TK: table caption goes on top as per LNCS style guide}

Table~\ref{tab:dmcs} shows some experimental results on calculating equilibria w.r.t. a randomly selected starting context of MSC with $n$ contexts, where $n$ varies between 9 and 151. Each context has an alphabet of $10$ atoms, exports at most $5$ atoms to other contexts, and has a maximum of 5 bridge rules with at most 2 bridge literals.
First, we confirm the results of Bairakdar \emph{et al.}~\cite{badaeifikr10a}, i.e., \systemname{dmcsopt} can handle larger sizes of MCSs more efficiently than \systemname{dmcs}.
Second, evaluating the MCS instances with symmetry breaking compared to the direct application of either \systemname{dmcs} or \systemname{dmcsopt} yields
improvements in response time throughout all tested topologies. In fact, symmetry breaking always leads to better runtimes, and in some cases, returns solutions to problems which are otherwise intractable within the given time.

\begin{figure}[t]
\centering
\includegraphics[width=.85\linewidth]{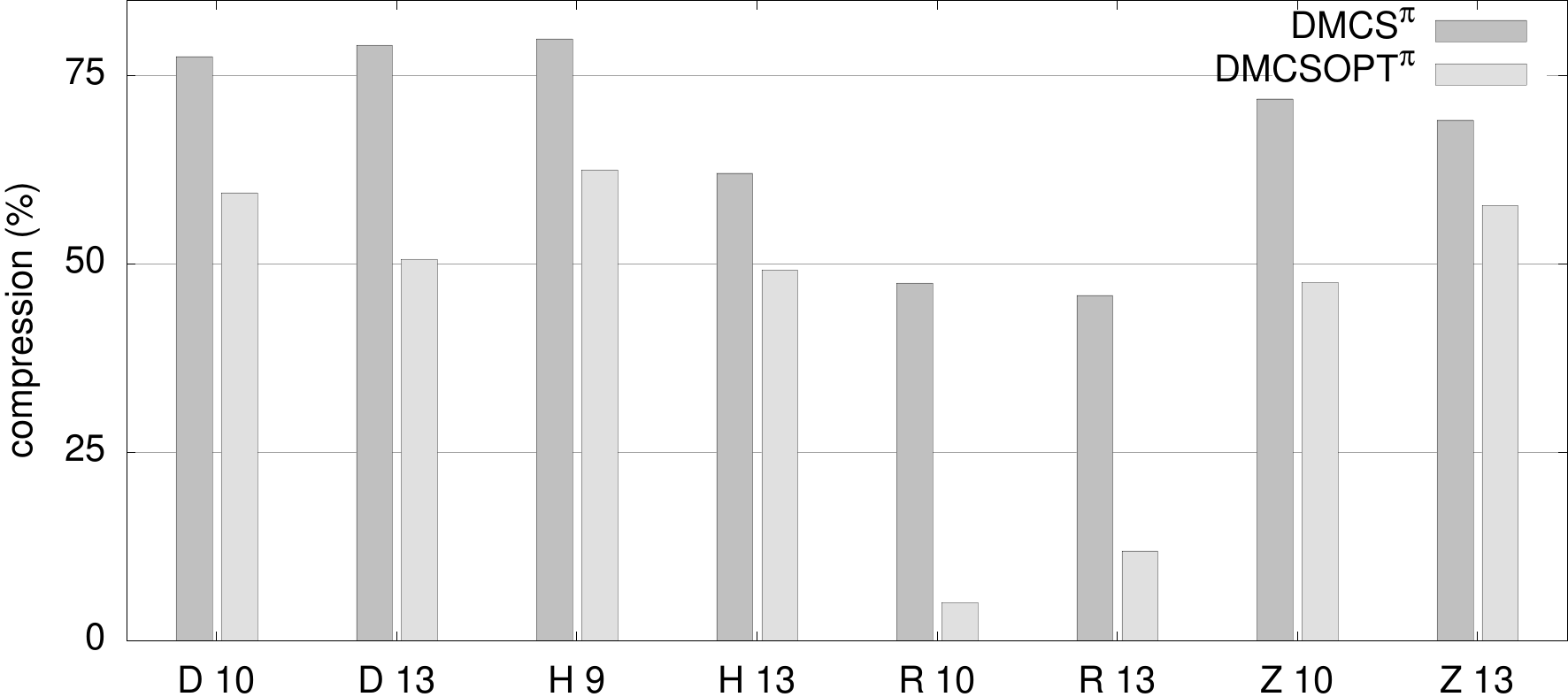}
\vspace{-1.5em}
\caption{
 Avg. compression of the solution space
 using local symmetry breaking
w.
 irred. 
  generators. \label{fig:dmcs} }
\vspace{-2.5em}
\end{figure}

Fig.~\ref{fig:dmcs}\comment{C: reduce chart' size? TK: done} presents the
average compression of the solution space achieved by symmetry
breaking. While the results for \systemname{dmcs}$^\pi$ range between 45\%
and 80\%,
the impact of symmetry breaking within
\systemname{dmcsopt} on the number of solutions varies between~5\% and~65\%.
We explain the latter with the restriction of \systemname{dmcsopt} to relevant atoms defined by the calling context.

\section{Conclusion}
We have presented a method for distributed symmetry detection and 
breaking for MCS.
In particular, we have designed a distributed algorithm such that each context computes its own (partial) symmetries and communicates them with another for exchanging partial symmetries in order to compute symmetries of the system as a whole. Distributed symmetry-breaking constraints prevent an evaluation engine from ever visiting two points in the search space that are equivalent under the symmetry they represent.
%
%
We have instantiated symmetry detection and symmetry breaking for MCS with ASP contexts, i.e., we have reduced partial symmetry of an ASP context to the automorphism of a coloured graph and encode symmetry breaking constraints as a distributed logic program.
Experiments on recent MCS benchmarks and show promising results.
Future work concerns a join operator for partial symmetries that preserves irredundant generators.

\iffinal\inlinereftrue\fi
\ifinlineref

\else
\bibliographystyle{splncs03}
\bibliography{paper}
\fi

\end{document}